\documentclass[review]{elsarticle}

\usepackage{lineno,hyperref}

\begin{document}

\begin{frontmatter}

\title{Gluon field fluctuations in nuclear collisions: Multiplicity and eccentricity distributions}


\author[BNL]{Bj\"orn Schenke}
\author[Tribedy]{Prithwish Tribedy}
\author[BNL]{Raju Venugopalan}

\address[BNL]{Physics Department, Brookhaven National Laboratory, Upton, NY 11973, USA}
\address[Tribedy]{Variable Energy Cyclotron Centre, 1/AF Bidhan Nagar, Kolkata 700064, India}



\begin{abstract}
We discuss different sources of fluctuations in nuclear collisions and their realization in the IP-Glasma model.
We present results for multiplicity distributions in p+p and p+A collisions and compare eccentricity ($\varepsilon_2$, $\varepsilon_3$, $\varepsilon_4$)
distributions in A+A collisions to the $v_n$ distributions in 10 centrality classes measured by the ATLAS collaboration.
\end{abstract}


\end{frontmatter}


\section{Introduction}

The Color Glass Condensate (CGC) effective field theory \cite{Gelis:2010nm} describes the generation of a dynamical transverse momentum scale $Q_s \gg \Lambda_{\rm QCD}$ at high energies, below which gluon fields saturate at occupancies of order $1/\alpha_s$. 
If the coupling runs as a function of this dynamical scale, $\alpha_s(Q_s)\ll 1$, weak coupling methods can be used to compute quantities that were believed previously to be intractable. This includes gluon production at low momenta where conventional perturbative methods would fail.

In this paper we use the CGC based IP-Glasma model \cite{Schenke:2012wb,Schenke:2012hg} to compute multiplicity distributions in proton-proton, proton-lead, and lead-lead collisions, as well as the initial shape of the collision, which we characterize by its eccentricities.

\section{IP-Glasma model}
The IP-Glasma model is described in detail in \cite{Schenke:2012wb,Schenke:2012hg}. It combines the IP-Sat model \cite{Kowalski:2003hm}, which determines $Q_s$ of a hadron or nucleus as a function of the gluon longitudinal momentum fraction $x$ and the transverse spatial position in the hadron or nucleus. The parameters of the model are determined by fits to deeply inelastic scattering data from HERA \cite{Rezaeian:2012ji}.

The IP-Glasma model samples nucleon positions for nuclei and uses the IP-Sat $Q_s$ distribution to compute the color charge density in the incoming hadrons/nuclei before the collision. It then samples individual color charges from this distribution. These moving color charges constitute the currents entering the classical Yang-Mills equations which determine the gluon fields within the fast moving hadrons/nuclei.

Finally, the fields at the moment of the collision are determined and evolved forward in time by means of the Yang-Mills equations. This provides the field energy-momentum tensor which can be used to generate initial conditions for hydrodynamic simulations \cite{Gale:2012rq}. One can further obtain gluon multiplicity distributions, which we will present in Section \ref{sec:mult}.
In Section \ref{sec:ecc} we discuss how to extract the spatial energy density distribution from $T^{\mu\nu}$, determine its eccentricities $\varepsilon_n$, and compare their distributions with experimental data for distributions of anisotropic flow coefficients $v_n$.

\section{Multiplicity distributions}\label{sec:mult}
\begin{figure}[t]
  \centering
  \includegraphics[width=10cm]{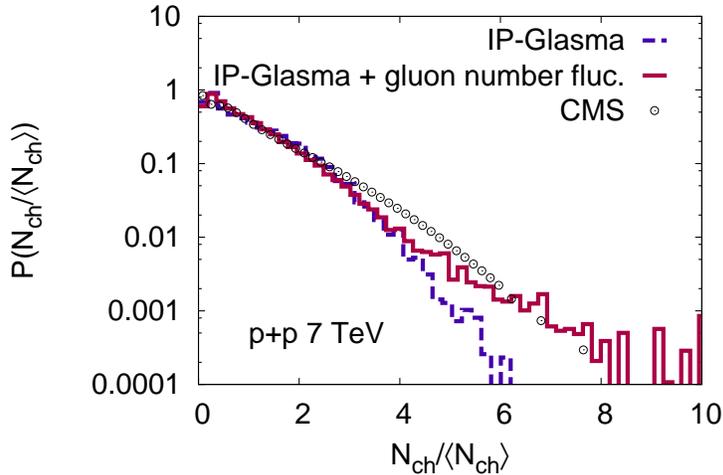} \vspace{-0.4cm}
  \caption{Charged hadron multiplicity distribution in $7\,{\rm TeV}$ p+p collisions compared to experimental data by the CMS collaboration \cite{Khachatryan:2010nk}. Result with and without additional fluctuations from fluctuating gluon numbers per flux tube.\label{fig:dNdy-pp} } 
\end{figure}

In this section we present results for multiplicity distributions in p+p and p+Pb collisions. In \cite{Schenke:2013dpa} we discussed how fluctuations of gluon numbers in a flux tube, that were neglected in the first implementation of the IP-Glasma model, can widen the distribution and achieve better agreement with the experimental data. There should be additional fluctuations in the hadronization process, leading to varying numbers of hadrons from a given number of gluons, also widening the theoretical multiplicity distribution. In Fig.\,\ref{fig:dNdy-pp} we show the result for the gluon multiplicity distribution with and without Gaussian fluctuations in the gluon number and compare to the experimental charged hadron multiplicity obtained by the CMS collaboration \cite{Khachatryan:2010nk}. The Gaussian gluon number fluctuations widen the distribution but do not reproduce its shape exactly. A modified distribution for these number fluctuations, inclusion of the NLO jet graph for gluon production and additional fluctuations from hadronization processes will modify the shown result and potentially lead to better agreement with the experimental data.

\begin{figure}[ht]
  \centering
  \includegraphics[width=10cm]{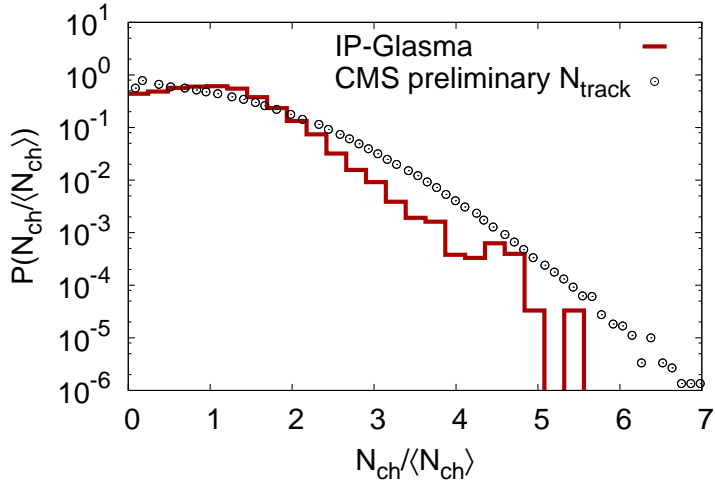} \vspace{-0.4cm}
  \caption{Charged hadron multiplicity distribution in $5.02\,{\rm TeV}$ p+Pb collisions compared to uncorrected experimental data for $N_{\rm track}$ by the CMS collaboration \cite{CMS:2012qk,Chatrchyan:2013nka}. Result with additional fluctuations from fluctuating gluon numbers per flux tube.\label{fig:dNdy-pPb} } 
\end{figure}

The same arguments hold for the multiplicity distribution in p+Pb collisions that are shown in Fig.\,\ref{fig:dNdy-pPb}, where we compare to uncorrected preliminary experimental data by the CMS collaboration \cite{CMS:2012qk,Chatrchyan:2013nka}. 

\section{Eccentricity distributions in Pb+Pb collisions}\label{sec:ecc}
\begin{figure}[t]
  \centering
  \begin{minipage}[b]{1.\linewidth}
    \includegraphics[width=6cm]{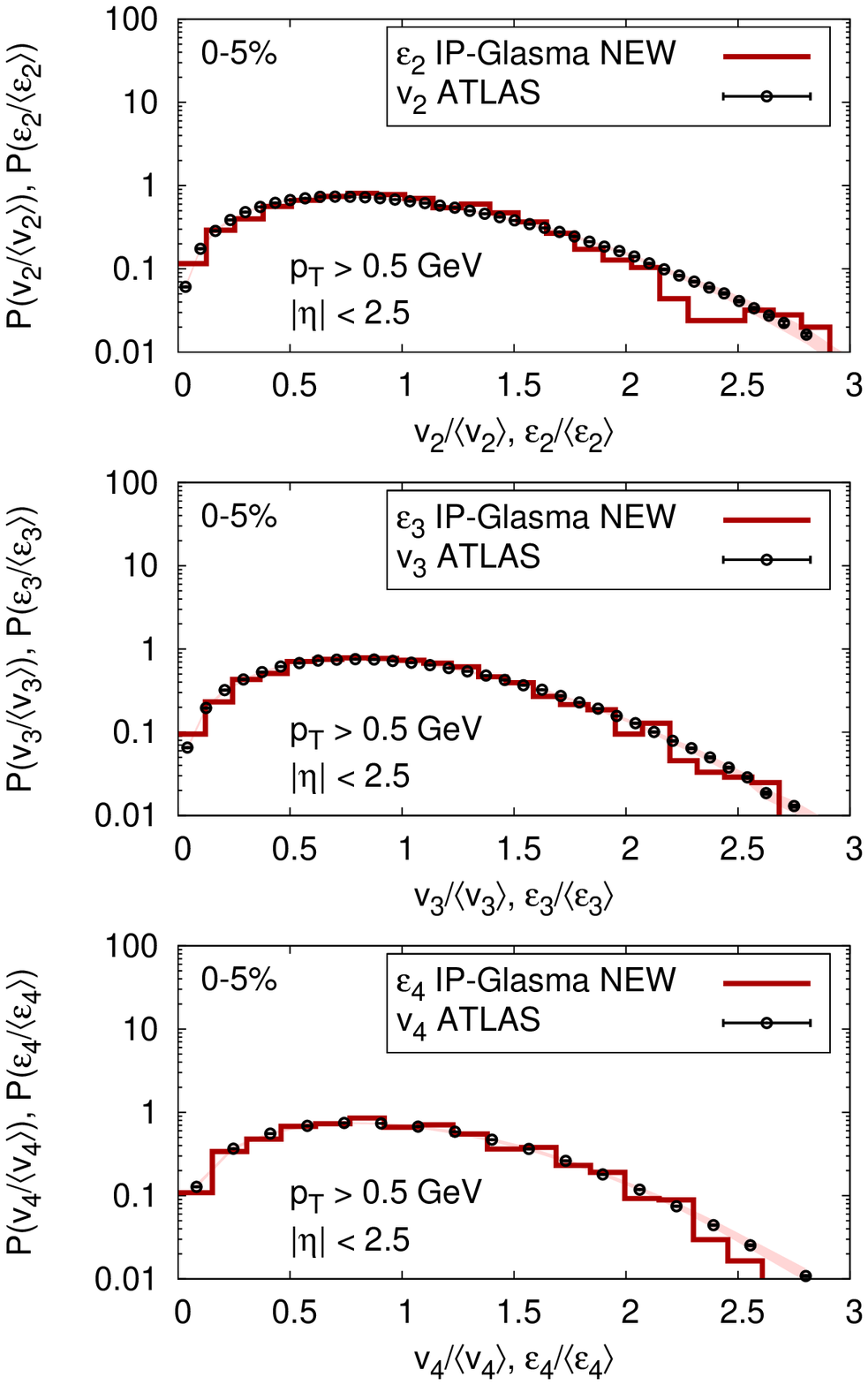}    \includegraphics[width=6cm]{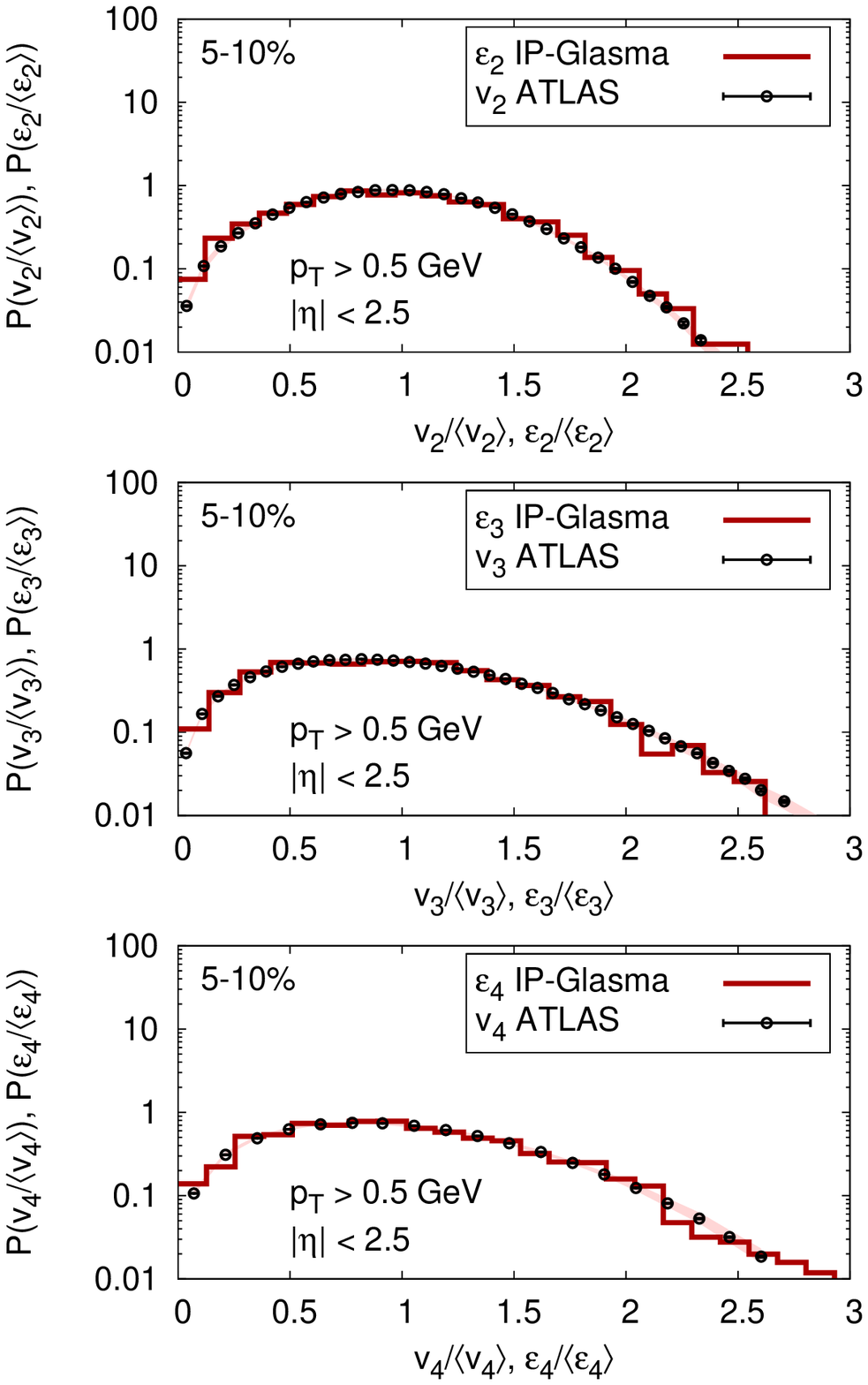}
  \end{minipage}
    \caption{Eccentricity $\varepsilon_n$ distributions from the IP-Glasma model for $n\in\{2,3,4\}$ compared to flow harmonic $v_n$ distributions measured by the ATLAS collaboration \cite{Aad:2013xma}. \label{fig:vnenDist1} } 
  \end{figure}

\begin{figure}[ht]
  \centering
  \begin{minipage}[b]{1.\linewidth}
    \includegraphics[width=6cm]{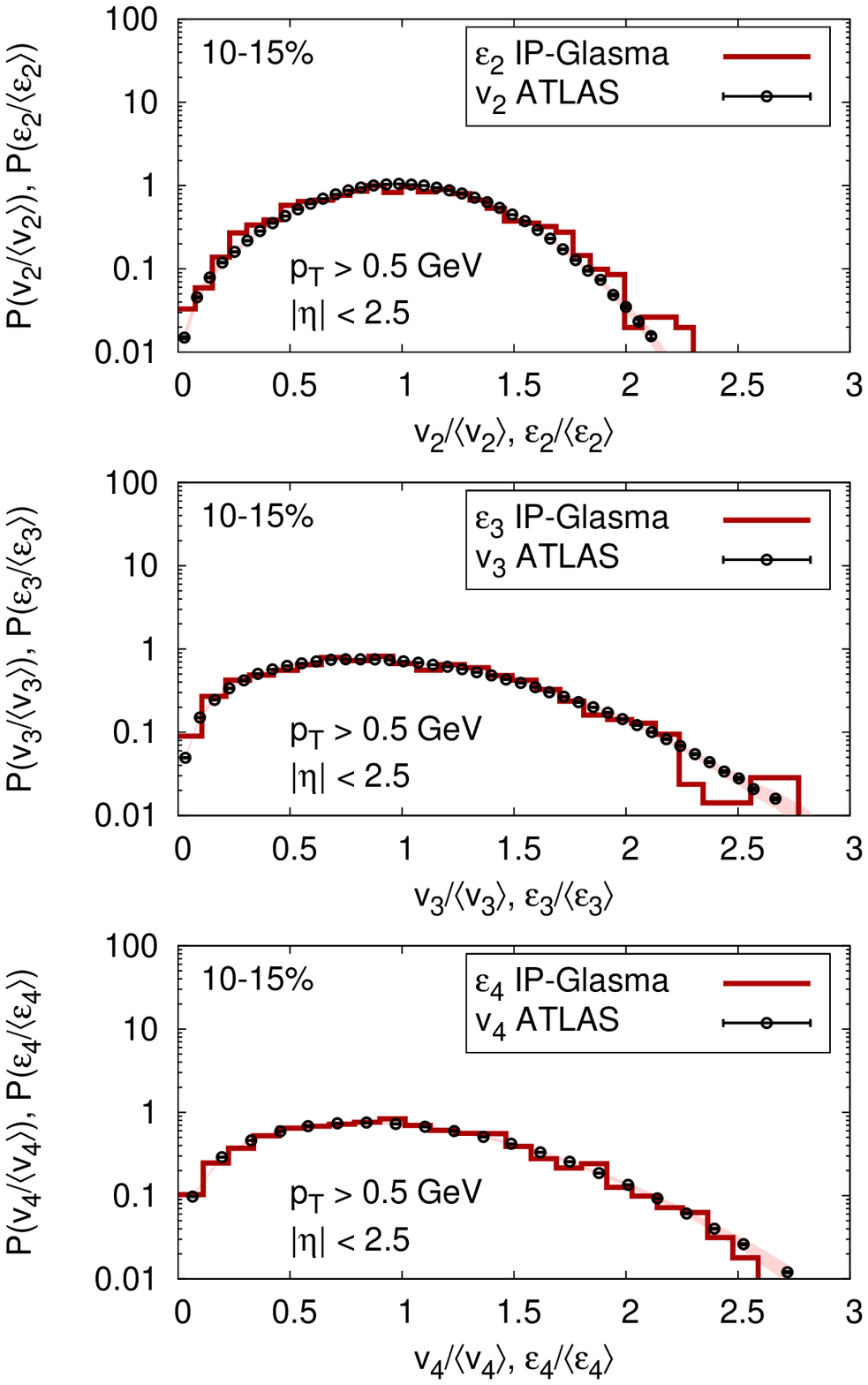}    \includegraphics[width=6cm]{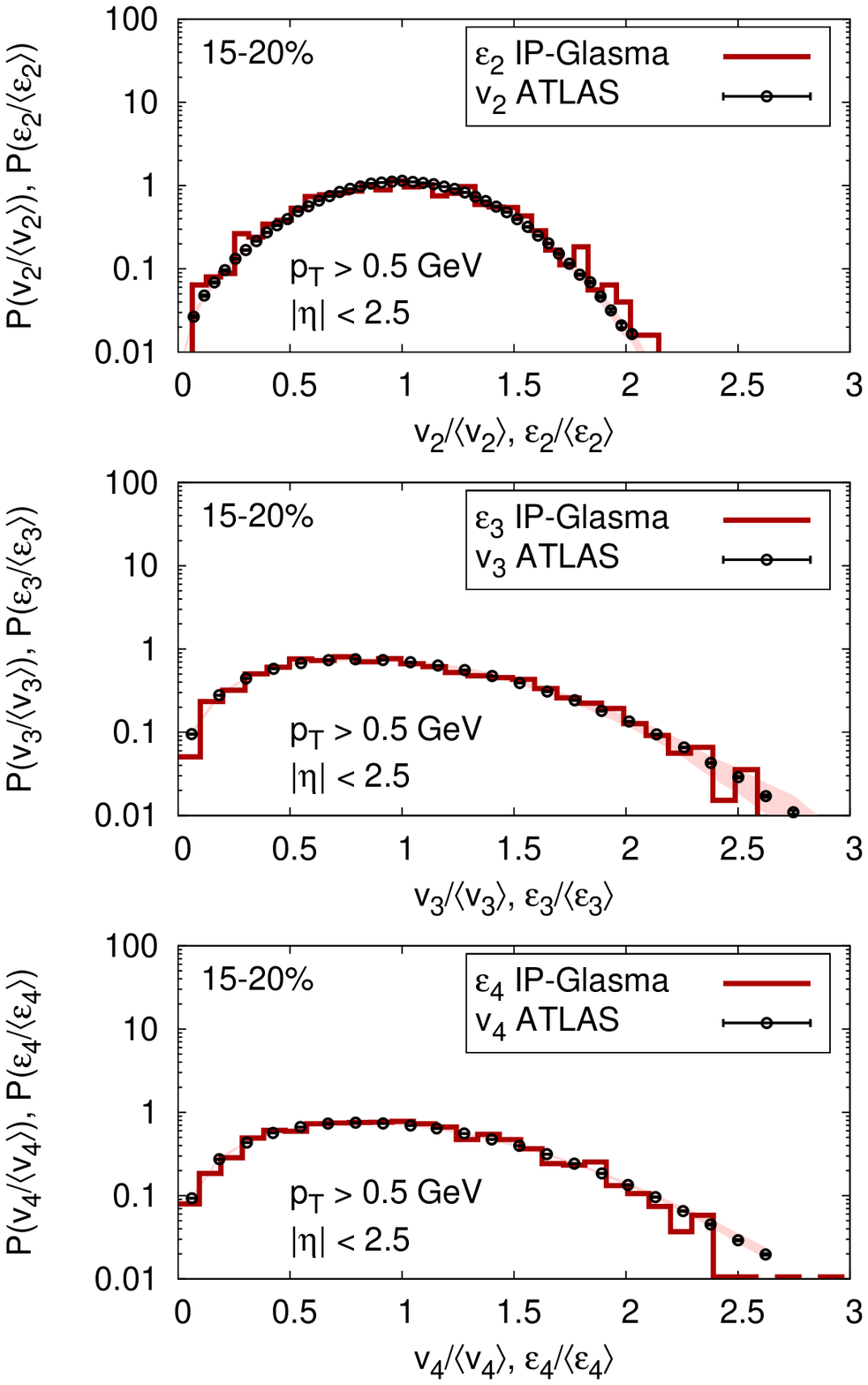}
  \end{minipage}
    \caption{Eccentricity $\varepsilon_n$ distributions from the IP-Glasma model for $n\in\{2,3,4\}$ compared to flow harmonic $v_n$ distributions measured by the ATLAS collaboration \cite{Aad:2013xma}. \label{fig:vnenDist2} } 
  \end{figure}

\begin{figure}[ht]
  \centering
  \begin{minipage}[b]{1.\linewidth}
    \includegraphics[width=6cm]{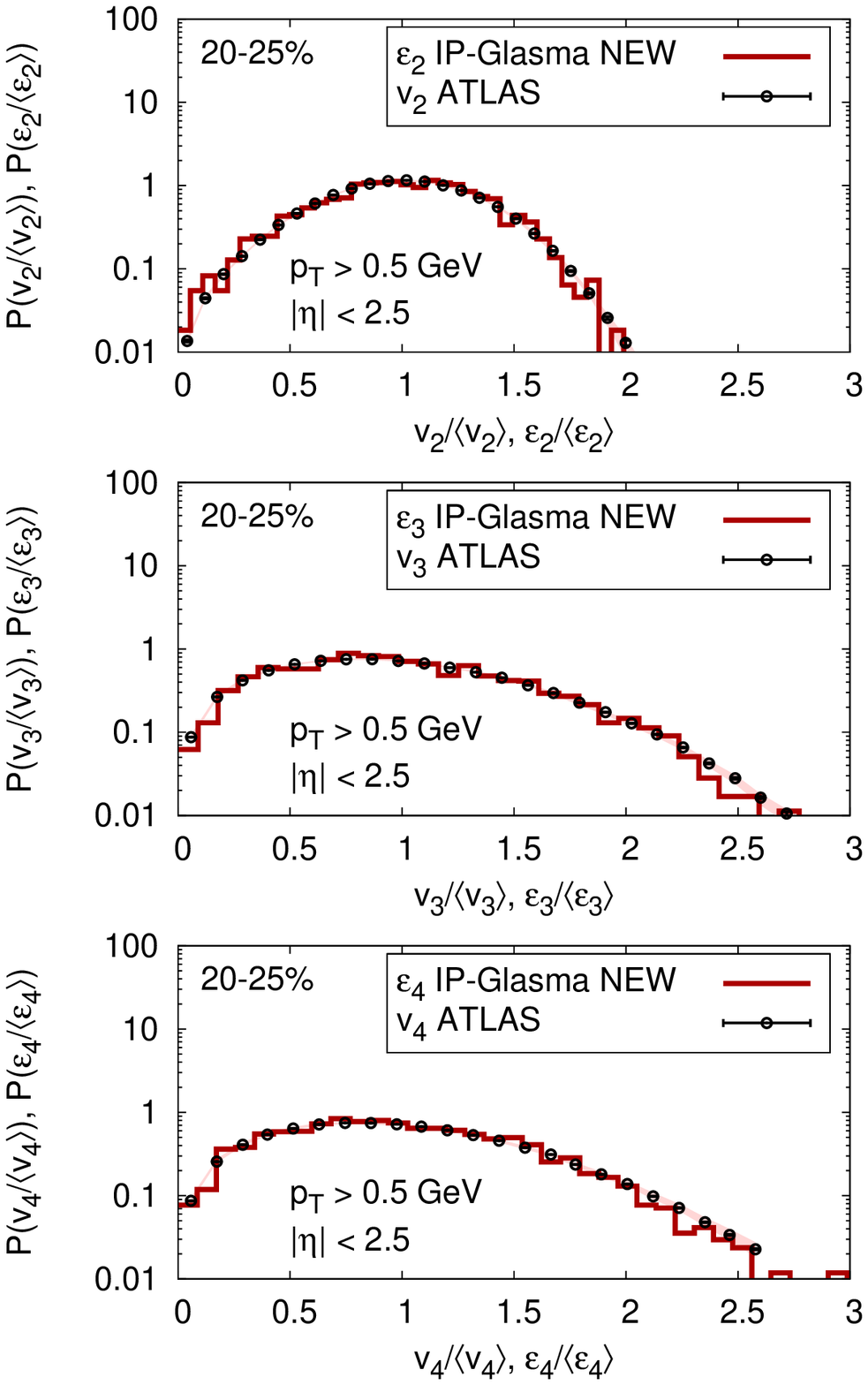}    \includegraphics[width=6cm]{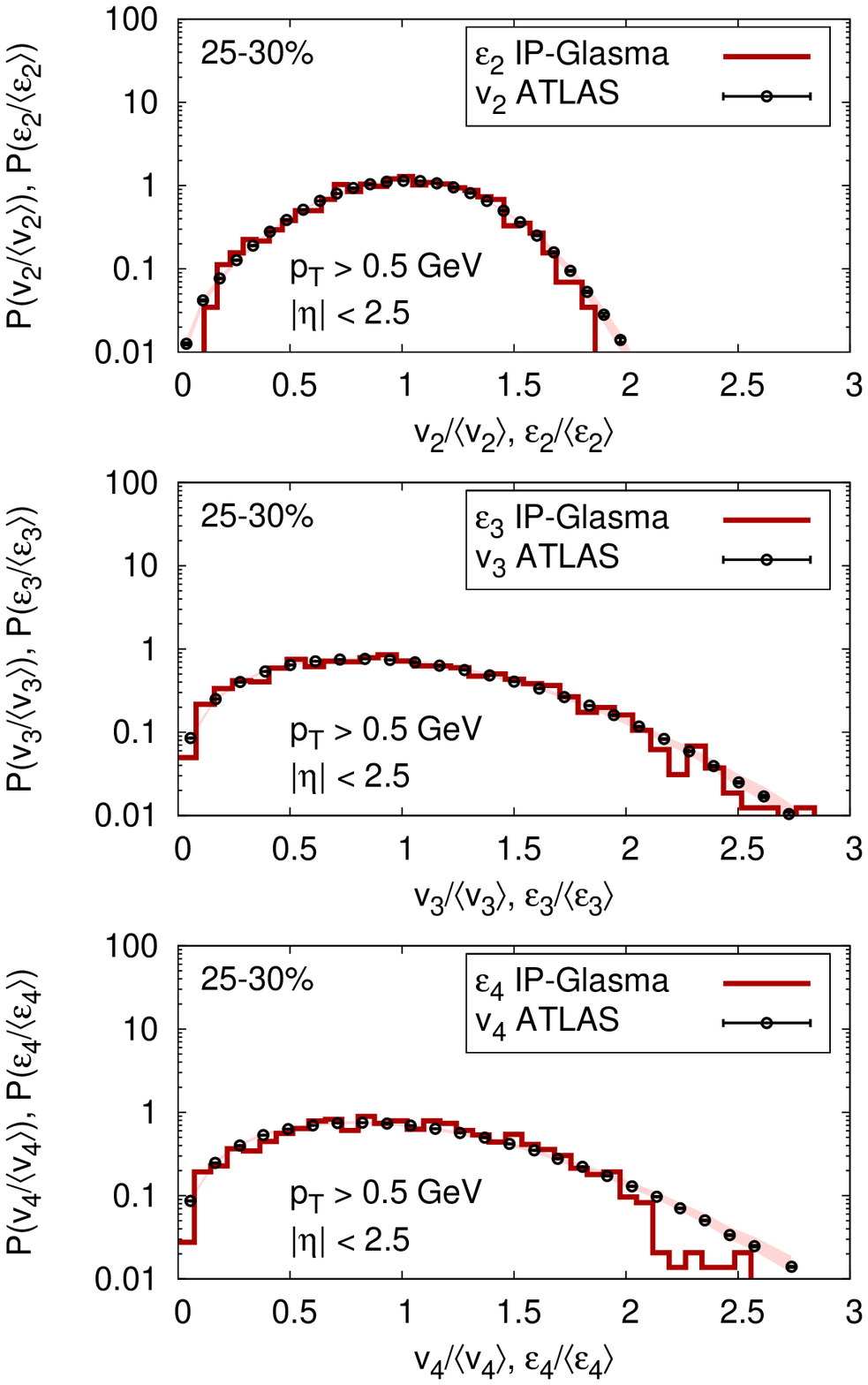}
  \end{minipage}
  \caption{Eccentricity $\varepsilon_n$ distributions from the IP-Glasma model for $n\in\{2,3,4\}$ compared to flow harmonic $v_n$ distributions measured by the ATLAS collaboration \cite{Aad:2013xma}.  \label{fig:vnenDist3}} 
\end{figure}

\begin{figure}[ht]
  \vspace{-0.4cm}
  \centering
  \begin{minipage}[b]{1.\linewidth}
    \includegraphics[width=6cm]{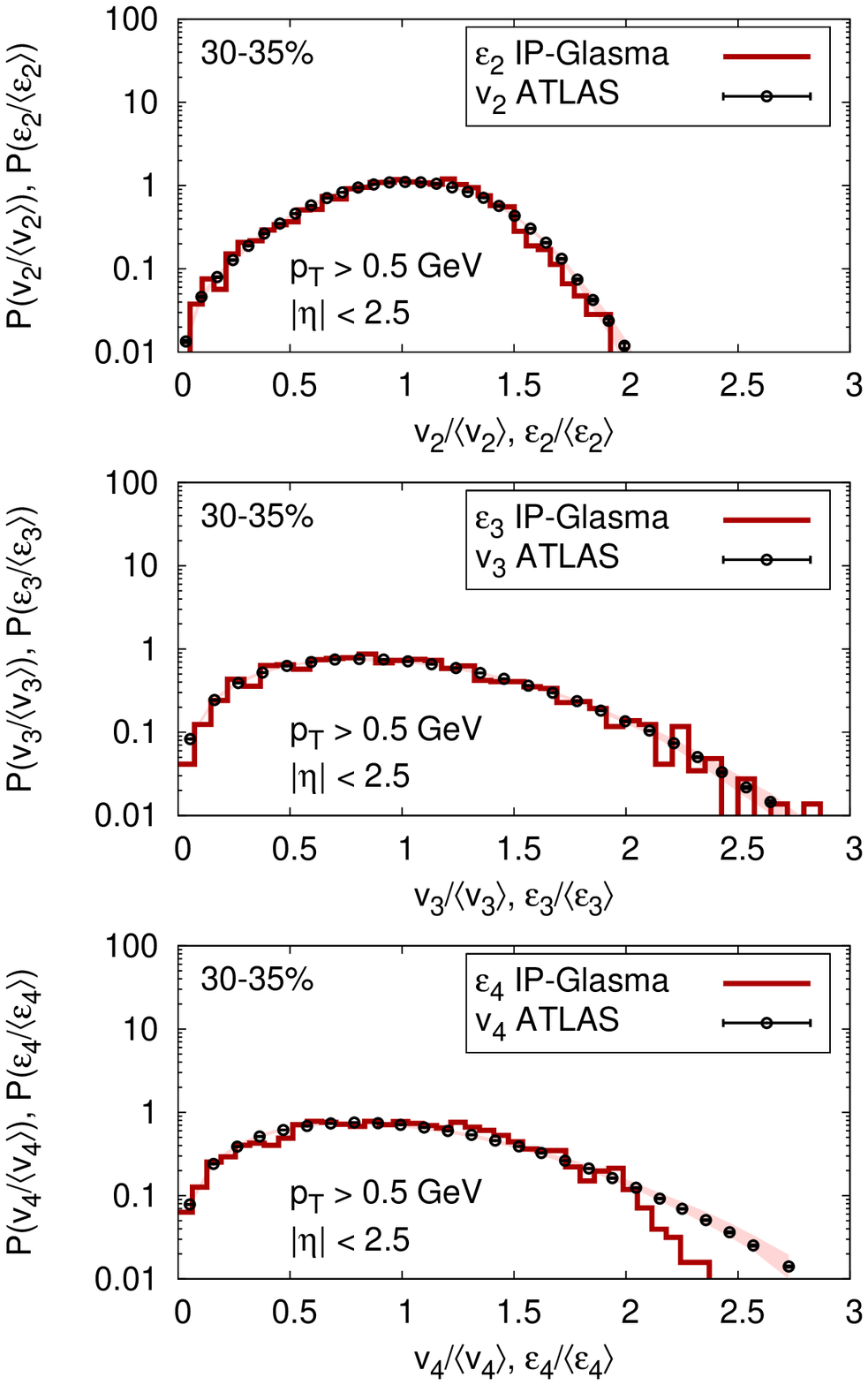}    \includegraphics[width=6cm]{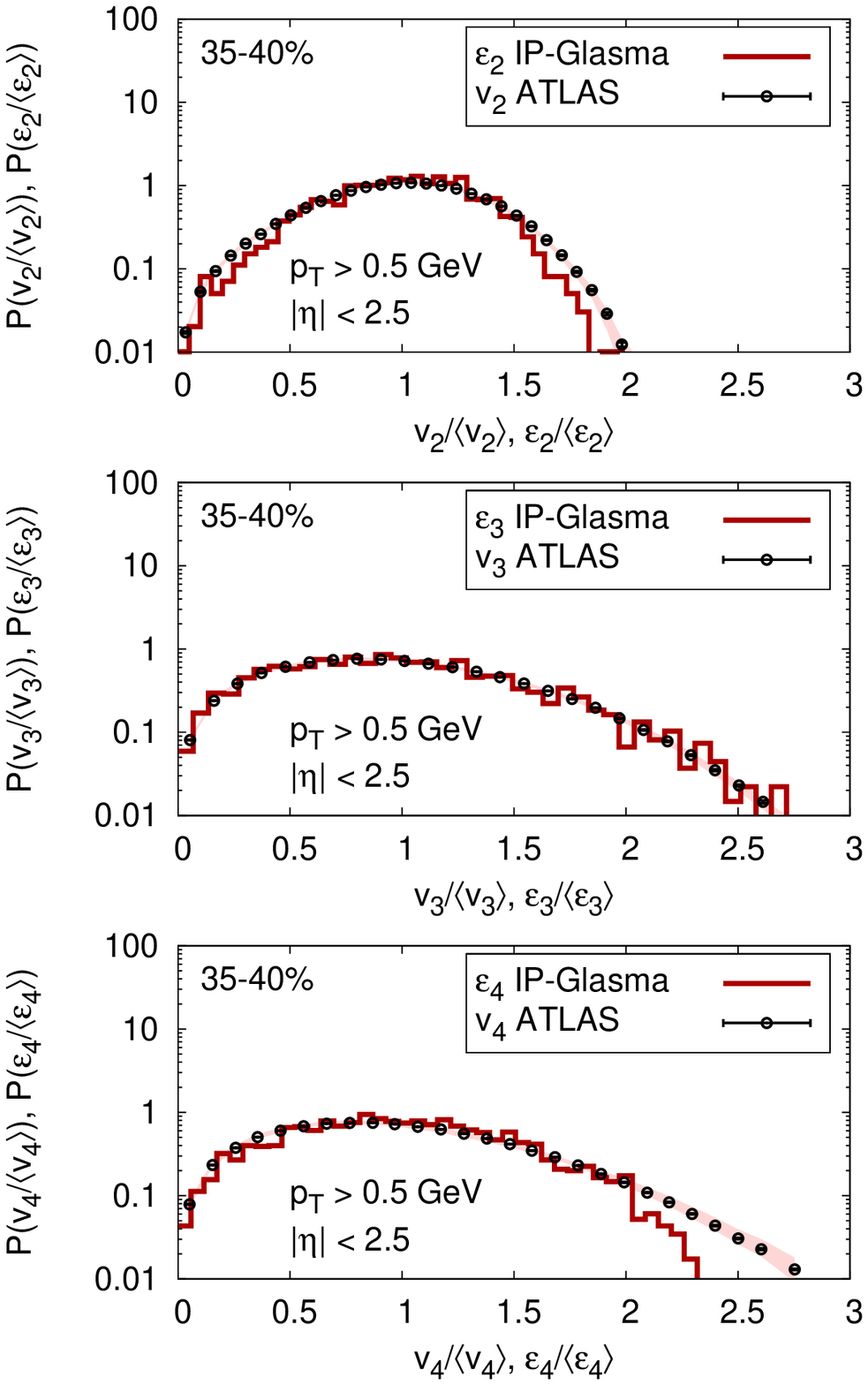}
  \end{minipage}
   \quad
  \begin{minipage}[b]{1.\linewidth}
    \includegraphics[width=6cm]{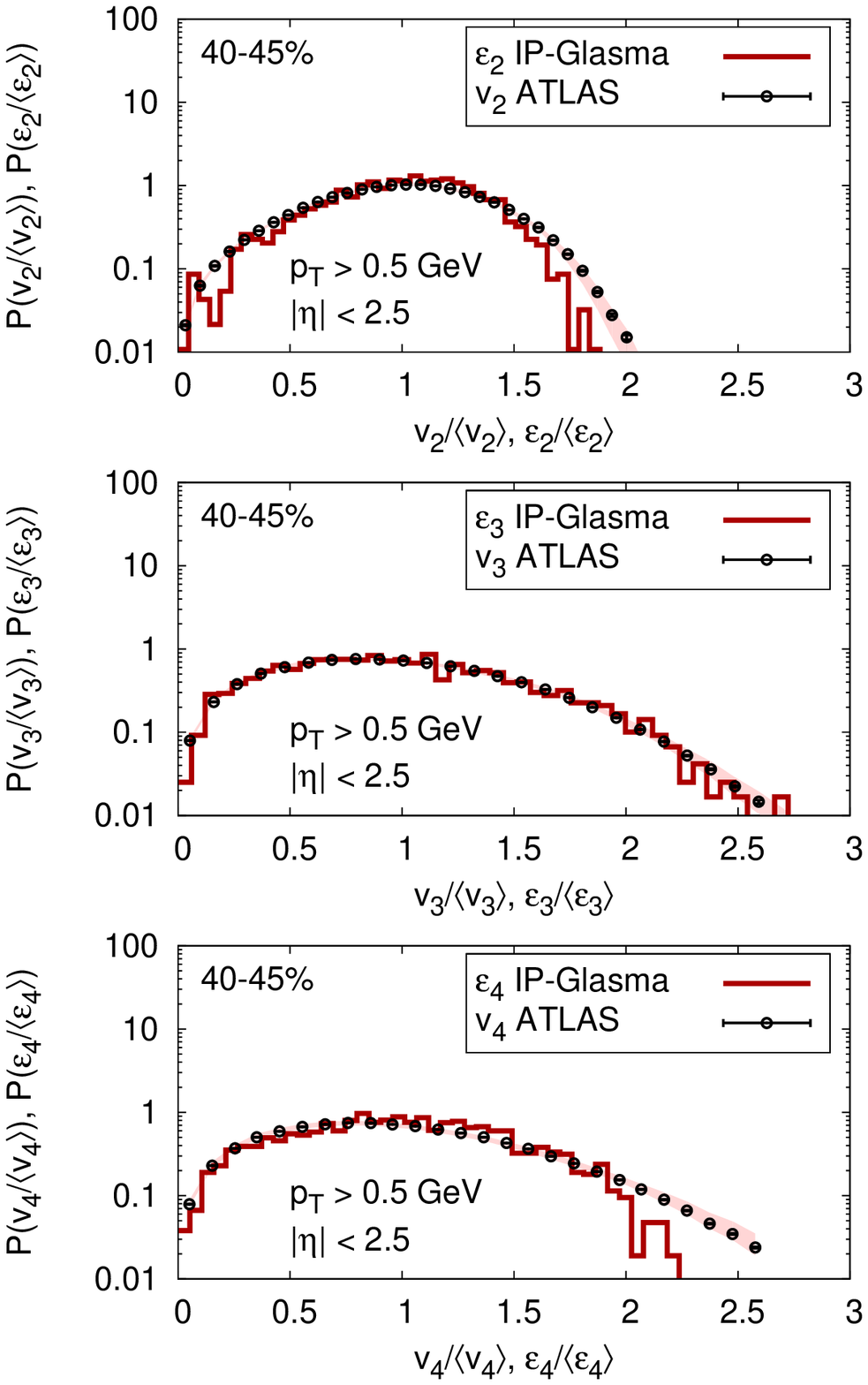}    \includegraphics[width=6cm]{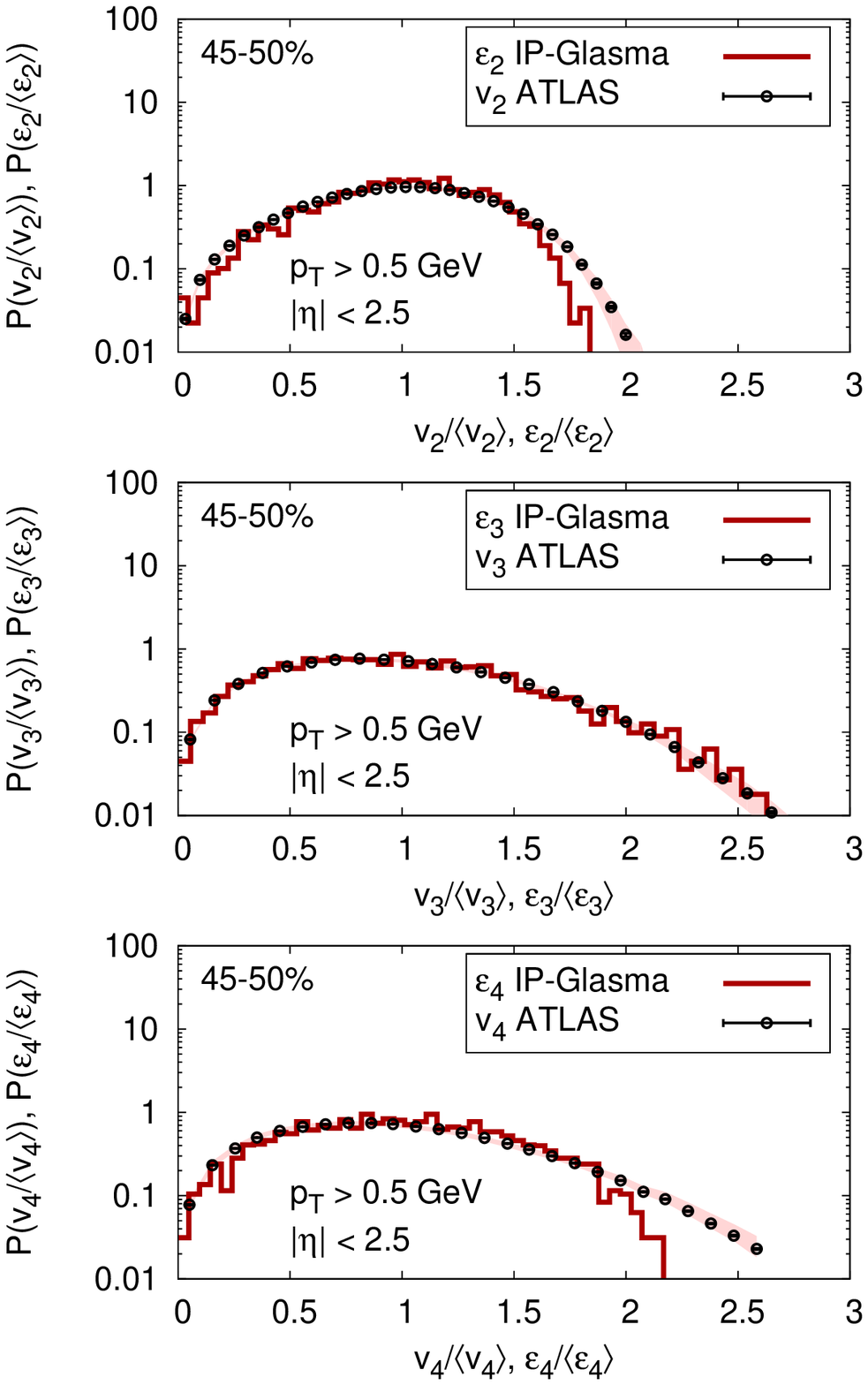}
  \end{minipage}\vspace{-0.3cm}
  \caption{Eccentricity $\varepsilon_n$ distributions from the IP-Glasma model for $n\in\{2,3,4\}$ compared to flow harmonic $v_n$ distributions measured by the ATLAS collaboration \cite{Aad:2013xma}.  \label{fig:vnenDist4}} 
\end{figure}

Next, we present event-by-event eccentricity distributions in $2.76\,{\rm TeV}$ Pb+Pb collisions for 10 centrality bins, comparing to experimental data for flow harmonics distributions by the ATLAS collaboration \cite{Aad:2013xma}. Results are shown in Figs.\,\ref{fig:vnenDist1} to \ref{fig:vnenDist4}. Eccentricities $\varepsilon_n = \sqrt{\langle r^n \cos(n\phi)\rangle^2+\langle r^n \sin(n\phi)\rangle^2}/\langle r^n \rangle$ are determined by using the energy density $\varepsilon$ as the weight in the spatial average $\langle \cdot \rangle$. The energy density follows from solving $u_\mu T^{\mu\nu} = \varepsilon u^\nu$, where $u^\mu$ are the fluid flow velocities and $T^{\mu\nu}$ is the field energy momentum tensor.

The eccentricities $\varepsilon_n$ are good predictors of the harmonic flow coefficients $v_n$ up to corrections from non-linear hydrodynamic evolution \cite{Gale:2012rq,Niemi:2012aj}. This is why their distributions, after dividing by the mean value $\langle \varepsilon_n \rangle$, provide a good description of the $v_n/\langle v_n\rangle$ distributions.
The non-linear corrections are important for the description of the large $v_n$ tail of the distributions \cite{Gale:2012rq}. 
This can be seen in the plots of $v_2$ and $v_4$ for centralities $>25\%$.
The reason is that the eccentricities miss the contribution to $v_4$ from non-linear coupling to the second harmonic and vice versa.
Certain events with a large $\varepsilon_2$ will generate a large contribution to $v_4$ during the here neglected evolution. In the large $v_4$ tail of the distribution even few such large $\varepsilon_2$ events will make a significant difference.

\section*{Acknowledgments}
BPS\ and RV\ are supported under DOE Contract No. DE-AC02-98CH10886. 
This research used resources of the National Energy Research Scientific Computing Center, supported 
by the Office of Science of the U.S. Department of Energy under Contract No. DE-AC02-05CH11231, and  
the Guillimin cluster at the CLUMEQ HPC centre, a part of Compute Canada HPC facilities.

\section*{References}

\bibliography{spires}

\begin{thebibliography}{12}
\expandafter\ifx\csname natexlab\endcsname\relax\def\natexlab#1{#1}\fi
\providecommand{\url}[1]{\texttt{#1}}
\providecommand{\href}[2]{#2}
\providecommand{\path}[1]{#1}
\providecommand{\DOIprefix}{doi:}
\providecommand{\ArXivprefix}{arXiv:}
\providecommand{\URLprefix}{URL: }
\providecommand{\Pubmedprefix}{pmid:}
\providecommand{\doi}[1]{\href{http://dx.doi.org/#1}{\path{#1}}}
\providecommand{\Pubmed}[1]{\href{pmid:#1}{\path{#1}}}
\providecommand{\bibinfo}[2]{#2}
\ifx\xfnm\relax \def\xfnm[#1]{\unskip,\space#1}\fi
\bibitem[{Gelis et~al.(2010)Gelis, Iancu, Jalilian-Marian, and
  Venugopalan}]{Gelis:2010nm}
\bibinfo{author}{F.~Gelis}, \bibinfo{author}{E.~Iancu},
  \bibinfo{author}{J.~Jalilian-Marian}, \bibinfo{author}{R.~Venugopalan},
  \bibinfo{journal}{Ann.Rev.Nucl.Part.Sci.} \bibinfo{volume}{60}
  (\bibinfo{year}{2010}) \bibinfo{pages}{463--489}.
  \DOIprefix\doi{10.1146/annurev.nucl.010909.083629}.
\bibitem[{Schenke et~al.(2012{\natexlab{a}})Schenke, Tribedy, and
  Venugopalan}]{Schenke:2012wb}
\bibinfo{author}{B.~Schenke}, \bibinfo{author}{P.~Tribedy},
  \bibinfo{author}{R.~Venugopalan}, \bibinfo{journal}{Phys. Rev. Lett.}
  \bibinfo{volume}{108} (\bibinfo{year}{2012}{\natexlab{a}})
  \bibinfo{pages}{252301}.
\bibitem[{Schenke et~al.(2012{\natexlab{b}})Schenke, Tribedy, and
  Venugopalan}]{Schenke:2012hg}
\bibinfo{author}{B.~Schenke}, \bibinfo{author}{P.~Tribedy},
  \bibinfo{author}{R.~Venugopalan}, \bibinfo{journal}{Phys. Rev.}
  \bibinfo{volume}{C86} (\bibinfo{year}{2012}{\natexlab{b}})
  \bibinfo{pages}{034908}.
\bibitem[{Kowalski and Teaney(2003)}]{Kowalski:2003hm}
\bibinfo{author}{H.~Kowalski}, \bibinfo{author}{D.~Teaney},
  \bibinfo{journal}{Phys. Rev.} \bibinfo{volume}{D68} (\bibinfo{year}{2003})
  \bibinfo{pages}{114005}. \DOIprefix\doi{10.1103/PhysRevD.68.114005}.
\bibitem[{Rezaeian et~al.(2013)Rezaeian, Siddikov, Van~de Klundert, and
  Venugopalan}]{Rezaeian:2012ji}
\bibinfo{author}{A.~H. Rezaeian}, \bibinfo{author}{M.~Siddikov},
  \bibinfo{author}{M.~Van~de Klundert}, \bibinfo{author}{R.~Venugopalan},
  \bibinfo{journal}{Phys.Rev.} \bibinfo{volume}{D87} (\bibinfo{year}{2013})
  \bibinfo{pages}{034002}. \href{http://arxiv.org/abs/1212.2974}{\tt
  arXiv:1212.2974}.
\bibitem[{Gale et~al.(2013)Gale, Jeon, Schenke, Tribedy, and
  Venugopalan}]{Gale:2012rq}
\bibinfo{author}{C.~Gale}, \bibinfo{author}{S.~Jeon},
  \bibinfo{author}{B.~Schenke}, \bibinfo{author}{P.~Tribedy},
  \bibinfo{author}{R.~Venugopalan}, \bibinfo{journal}{Phys.Rev.Lett.}
  \bibinfo{volume}{110} (\bibinfo{year}{2013}) \bibinfo{pages}{012302}.
  \DOIprefix\doi{10.1103/PhysRevLett.110.012302}.
  \href{http://arxiv.org/abs/1209.6330}{\tt arXiv:1209.6330}.
\bibitem[{Schenke et~al.(2013)Schenke, Tribedy, and
  Venugopalan}]{Schenke:2013dpa}
\bibinfo{author}{B.~Schenke}, \bibinfo{author}{P.~Tribedy},
  \bibinfo{author}{R.~Venugopalan}  (\bibinfo{year}{2013}).
  \href{http://arxiv.org/abs/1311.3636}{\tt arXiv:1311.3636}.
\bibitem[{Khachatryan et~al.(2011)}]{Khachatryan:2010nk}
\bibinfo{author}{V.~Khachatryan}, et~al. (\bibinfo{collaboration}{CMS
  Collaboration}), \bibinfo{journal}{JHEP} \bibinfo{volume}{1101}
  (\bibinfo{year}{2011}) \bibinfo{pages}{079}.
  \DOIprefix\doi{10.1007/JHEP01(2011)079}.
  \href{http://arxiv.org/abs/1011.5531}{\tt arXiv:1011.5531}.
\bibitem[{Chatrchyan et~al.(2013{\natexlab{a}})}]{CMS:2012qk}
\bibinfo{author}{S.~Chatrchyan}, et~al. (\bibinfo{collaboration}{CMS
  Collaboration}), \bibinfo{journal}{Phys.Lett.} \bibinfo{volume}{B718}
  (\bibinfo{year}{2013}{\natexlab{a}}) \bibinfo{pages}{795--814}.
  \DOIprefix\doi{10.1016/j.physletb.2012.11.025}.
  \href{http://arxiv.org/abs/1210.5482}{\tt arXiv:1210.5482}.
\bibitem[{Chatrchyan et~al.(2013{\natexlab{b}})}]{Chatrchyan:2013nka}
\bibinfo{author}{S.~Chatrchyan}, et~al. (\bibinfo{collaboration}{CMS
  Collaboration}), \bibinfo{journal}{Phys.Lett.} \bibinfo{volume}{B724}
  (\bibinfo{year}{2013}{\natexlab{b}}) \bibinfo{pages}{213--240}.
  \DOIprefix\doi{10.1016/j.physletb.2013.06.028}.
  \href{http://arxiv.org/abs/1305.0609}{\tt arXiv:1305.0609}.
\bibitem[{Aad et~al.(2013)}]{Aad:2013xma}
\bibinfo{author}{G.~Aad}, et~al. (\bibinfo{collaboration}{ATLAS
  Collaboration}), \bibinfo{journal}{JHEP} \bibinfo{volume}{1311}
  (\bibinfo{year}{2013}) \bibinfo{pages}{183}.
  \DOIprefix\doi{10.1007/JHEP11(2013)183}.
  \href{http://arxiv.org/abs/1305.2942}{\tt arXiv:1305.2942}.
\bibitem[{Niemi et~al.(2013)Niemi, Denicol, Holopainen, and
  Huovinen}]{Niemi:2012aj}
\bibinfo{author}{H.~Niemi}, \bibinfo{author}{G.~Denicol},
  \bibinfo{author}{H.~Holopainen}, \bibinfo{author}{P.~Huovinen},
  \bibinfo{journal}{Phys.Rev.} \bibinfo{volume}{C87} (\bibinfo{year}{2013})
  \bibinfo{pages}{054901}. \DOIprefix\doi{10.1103/PhysRevC.87.054901}.
  \href{http://arxiv.org/abs/1212.1008}{\tt arXiv:1212.1008}.

\end{thebibliography}

\end{document}